\documentclass[12pt]{article}
		\usepackage{graphicx}		% <for>inclusion of graphics</for>
		\usepackage{amsmath}
		\usepackage{amssymb}
		\usepackage{pict2e}		% <for>drawing pictures</for>
		\usepackage{ifthen}		% <for>conditionals</for>
		\usepackage[sort&compress]{natbib}		% <for>references</for>
		\usepackage{abbrvate}		% <for>abbreviations</for>
		\usepackage[usenames]{color}	% <for>color</for>
		\usepackage{bbm}			% <for>double struck letters (\mathbbm)</for>
		\newabbr{AdS}{anti-de Sitter}
		\newabbr{BG}{Batell-Gherghetta}
		\newabbr{BSM}{beyond the Standard Model}
		\newabbr{CFT}{conformal field theory}
		\newabbr{CKM}{Cabbibo-Kobayashi-Maskawa}
		\newabbr{EWSB}{electroweak symmetry breaking}
		\newabbr{GUT}{grand unified theory}
		\newabbr{IR}{infrared}
		\newabbr{KK}{Kaluza-Klein}
		\newabbr{LEP}{large electron positron collider}
		\newabbr{LHC}{large hadron collider}
		\newabbr{QCD}{quantum chromodynamics}
		\newabbr{RGE}{renormalization group equation}
		\newabbr{SM}{Standard Model}
		\newabbr{SSB}{spontaneous symmetry breaking}
		\newabbr{SUSY}{supersymmetry}
		\newabbr{UV}{ultra violet}
		\newabbr{VEV}{vacuum expectation value}
		\newabbr{WKB}{Wentzel-Kramers-Brillouin}
		\newcommand{\eq}[1]{Eq.~\eqref{Eq:#1}}
		\newcommand{\eqn}[1]{\eqref{Eq:#1}}
		\newcommand{\fig}[1]{Figure~\ref{Fig:#1}}
		\newcommand{\sect}[1]{Section~\ref{Sec:#1}}
		
		\newcommand{\app}[1]{Appendix~\ref{App:#1}}
			\newcommand{\EE}{\mathbbm{e}}
			
		\newcommand{\half}{\frac{1}{2}}			% 1/2
		\newcommand{\third}{\frac{1}{3}}			% 1/3
		\newcommand{\fourth}{\frac{1}{4}}		% 1/4
		\DeclareMathOperator{\Ei}{Ei}			% Exponential Integral Ei
			\newcommand{\gD}{\mathrm{D}}
			\newcommand{\pderiv}[3][]{%
				\ifthenelse{\equal{#1}{}}%
					{%
					\frac{ \partial #2}{ \partial #3 }%
					}%
					{%	
					\frac{ \partial^{#1} #2}{ \partial {#3}^{#1} }%
					}%
				}
			\newcommand{\pD}{\partial}
			\newcommand{\deriv}[3][]{%
				\ifthenelse{\equal{#1}{}}%
					{%
					\frac{ d #2}{ d #3 }%
					}%
					{%
					\frac{ d^{#1} #2}{ d {#3}^{#1} }%
					}%
				}
			\newcommand{\fderiv}[3][]{%
				\ifthenelse{\equal{#1}{}}%
					{%	
					\frac{ \delta #2}{ \delta {#3} }%
					}%
					{%
					\frac{ \delta^{#1} #2}{ \delta {#3}^{#1} }%
					}%
				}
			\newcommand{\intOp}[2][]{\int \! d^{#1}#2 \;}
		\newcommand{\abs}[1]{\mathopen{}\left|{#1}\right|}

		\DeclareMathOperator{\diagOp}{diag}
		\newcommand{\diag}[1]{\diagOp\inp{#1}}
		\newcommand{\inp}[2][0cm]{\mathopen{}\left(#2\parbox[h][#1]{0cm}{}\right)}
		\newcommand{\inb}[2][0cm]{\mathopen{}\left[#2\parbox[h][#1]{0cm}{}\right]}
		\newcommand{\inap}[2][0cm]{\mathopen{}\left<#2\parbox[h][#1]{0cm}{}\right>}
		\newcommand{\vev}[1]{\inap{#1}}

\begin{document}
\pagestyle{empty}

\begin{center}

{\Large\bf On the stability of a soft-wall model}

\vspace{1cm}

%
%	<authors>
			{\sc\normalsize T.~Gherghetta\footnote{\texttt{tgher@unimelb.edu.au}} \ 
			and
			N.~Setzer\footnote{\texttt{nsetzer@unimelb.edu.au}}} \\%
			\vspace{0.2cm}
		      \textit{\normalsize School of Physics, University of Melbourne, Victoria, 3010, Australia}%
%	</authors>
%
\end{center}

\vspace{2cm}

%%%%%%%%%%%%%%%%%%%%%%%%%%%%%%%%%%%%%%%%%%%%%%%%%%%%%%%%%%%%%%%%%%%%%%%%%%%%%%%%%%%%%%%%%%%%%%%%%%%%%%%%%%%%%%%%%%%%%
% * * * * * * * * * * * * * * * * * * * * * * * * * * * * * * * * * * * * * * * * * * * * * * * * * * * * * * * * * *
\abbrstyle{expandall}% always expand abbreviations in the abstract
\begin{abstract}
We investigate the stability and fluctuations of a soft wall model that has an asymptotically AdS metric and a scalar field that has an asymptotically power-law dependence in the conformal coordinate.  By imposing UV boundary conditions, the soft wall mass scale can be fixed to be near the TeV scale and causes the radion to no longer be massless.  A hierarchy between the weak scale and the Planck scale can be generated for various particle spectrum behavior, although natural values only occur for a gravitational sector containing scalar fields that act like unparticles.  In addition, if bulk Standard Model fields have nonstandard couplings to the gravitational sector, then a discrete particle spectrum can be realized in the nongravitational sector.  This allows for the possibility of an unparticle solution to the hierarchy problem.
\end{abstract}
% * * * * * * * * * * * * * * * * * * * * * * * * * * * * * * * * * * * * * * * * * * * * * * * * * * * * * * * * * *
%%%%%%%%%%%%%%%%%%%%%%%%%%%%%%%%%%%%%%%%%%%%%%%%%%%%%%%%%%%%%%%%%%%%%%%%%%%%%%%%%%%%%%%%%%%%%%%%%%%%%%%%%%%%%%%%%%%%%

% <abbreviations>
	\abbrstyle{plain} 	% use plain abbreviation style
	\abbrreset		% reset all abbreviations (in case used in abstract)
	\abbrmakeused[CKM]%	always use abbreviation for CKM
	\abbrmakeused[LEP]%	always use abbreviation for LEP
	\abbrmakeused[QCD]%	always use abbreviation for QCD		
	\abbrmakeused[WKB]%	always use abbreviation for WKB
	\abbrmakeused[UV]%	always use abbreviation for UV
% </abbreviations>

\newpage
\setcounter{page}{1}
\setcounter{footnote}{0}
\pagestyle{plain}

%%%%%%%%%%%%%%%%%%%%%%%%%%%%%%%%%%%%%%%%%%%%%%%%%%%%%%%%%%%%%%%%%%%%%%%%%%%%%%%%%%%%%%%%%%%%%%%%%%%%%%%%%%%%%%%%%%%%%
%%%%%%%%%%%%%%%%%%%%%%%%%%%%%%%%%%%%%%%%%%%%%%%%%%%%%%%%%%%%%%%%%%%%%%%%%%%%%%%%%%%%%%%%%%%%%%%%%%%%%%%%%%%%%%%%%%%%%
\section{Introduction}
\label{Sec:Intro}
%%%%%%%%%%%%%%%%%%%%%%%%%%%%%%%%%%%%%%%%%%%%%%%%%%%%%%%%%%%%%%%%%%%%%%%%%%%%%%%%%%%%%%%%%%%%%%%%%%%%%%%%%%%%%%%%%%%%%
%%%%%%%%%%%%%%%%%%%%%%%%%%%%%%%%%%%%%%%%%%%%%%%%%%%%%%%%%%%%%%%%%%%%%%%%%%%%%%%%%%%%%%%%%%%%%%%%%%%%%%%%%%%%%%%%%%%%%

An elegant way to solve the Planck-weak scale hierarchy problem is to use a warped extra dimension~\cite{Randall:1999ee}.  In addition to addressing this deficiency of the \abbr{SM}, a compact warped extra dimension can also be used to explain fermion mass hierarchies~\cite{Grossman:1999ra,Gherghetta:2000qt,Huber:2000ie}.  Another interesting feature of these $5D$ models is that while the warped dimension provides a weakly-coupled description, remarkably---via the correspondence between \abbr{AdS} space and \abbr{CFT}~\cite{Maldacena:1997re}---a four-dimensional ($4D$) strongly-coupled dual description can be given~\cite{ArkaniHamed:2000ds,Rattazzi:2000hs,PerezVictoria:2001pa}.

An example of this duality is seen in the usual compact warped extra dimension where an \abbr{IR} brane represents a ``hard-wall'' cutoff where the fifth dimension abruptly ends; in the $4D$ dual theory this corresponds to spontaneously breaking conformal symmetry at a particular \abbr{IR} scale.  It is possible, however, to slowly turn on the conformal breaking in the $4D$ theory by replacing the \abbr{IR} brane of the $5D$ theory with a dilaton field having a \abbr{VEV} dependent on the extra dimension.  This generalizes the idealized notion of a hard wall to a ``soft wall" and provides a more realistic framework of the possible conformal symmetry breaking dynamics.  In particular, the soft wall leads to a deviation in the spacing of the particle mass spectrum and was first proposed to obtain a linear Regge-like spectrum~\cite{Karch:2006pv}.

In the soft wall description, it is natural to ask whether this setup can describe electroweak physics in a manner similar to the hard wall scenarios.  Indeed, the soft wall can accommodate bulk \abbr{SM} fields propagating in the extra dimension~\cite{Falkowski:2008fz,Batell:2008me,Delgado:2009xb, MertAybat:2009mk, Gherghetta:2009qs,Sword:2009wt,Atkins:2010cc} and its stability can be addressed in a general class of models~\cite{Cabrer:2009we, Aybat:2010sn}.  In this work we shall address the stability and perform a fluctuation analysis of a specific solution related to that proposed in~\cite{Batell:2008zm,Batell:2008me}.  The solution is asymptotically \abbr{AdS} with a single scalar field that has an asymptotic power-law dependence in conformal coordinates with an arbitrary exponent.  This scalar \abbr{VEV} allows a description of all types of particle mass spectrum behavior such as discrete, continuum with a mass gap, and continuum without a mass gap to be obtained.  It is shown that the soft-wall background is stable and the radion is no longer massless.  Furthermore, we demonstrate that the introduction of \abbr{SM} fields on this background leads to phenomenological models where the Planck-weak scale hierarchy is stabilized.  For these stabilized models the gravitational sector can take on any type of particle mass spectra; however, natural values of the parameters only occur when the gravitational sector is a continuum or has unparticle behavior.  In this natural regime, it is necessary that the nongravitational sector containing the \abbr{SM} fields have nonstandard couplings of the dilaton so that they may obtain a mass spectrum of at least a continuum with a mass gap.  Due to this nonstandard coupling, a careful accounting of the backreaction is needed, and we present a condition on the scalar potential such that the backreaction of the \abbr{SM} fields can be neglected.  Together this suggests that there may be an unparticle solution to the hierarchy problem. 

The study presented here overlaps with the approach of~\cite{Cabrer:2009we} except that our geometry represents a different class of solutions.  Moreover, it contains a parameter that permits us to continuously vary the soft-wall model between an RS1-like and RS2-like limit.  The single scalar field is therefore simultaneously responsible for breaking conformal symmetry as well as stabilizing the brane configuration in a way that is similar to the Goldberger-Wise mechanism~\cite{Goldberger:1999uk,DeWolfe:1999cp}.

The paper is organized as follows: we begin in \sect{Soft-Wall.Model} by reviewing the background geometry and extracting the key features needed to address stability.  In \sect{Hierarchy} we explore the regime where a viable Planck-weak scale hierarchy is obtained, showing that the soft-wall term in the metric must contain the conformal coordinate to a fractional power.  The fluctuations around the background solutions are investigated in \sect{Fluctuations} and demonstrate that there are no 4D tachyons as well as no unwanted 4D zero mass modes.  In \sect{Bulk.Fields}, we address the issues of additional bulk gauge and scalar fields propagating in the soft-wall background, after which we conclude in \sect{Conclusion}.

%%%%%%%%%%%%%%%%%%%%%%%%%%%%%%%%%%%%%%%%%%%%%%%%%%%%%%%%%%%%%%%%%%%%%%%%%%%%%%%%%%%%%%%%%%%%%%%%%%%%%%%%%%%%%%%%%%%%%
%%%%%%%%%%%%%%%%%%%%%%%%%%%%%%%%%%%%%%%%%%%%%%%%%%%%%%%%%%%%%%%%%%%%%%%%%%%%%%%%%%%%%%%%%%%%%%%%%%%%%%%%%%%%%%%%%%%%%
\section{The Soft-Wall Model}
\label{Sec:Soft-Wall.Model}
%%%%%%%%%%%%%%%%%%%%%%%%%%%%%%%%%%%%%%%%%%%%%%%%%%%%%%%%%%%%%%%%%%%%%%%%%%%%%%%%%%%%%%%%%%%%%%%%%%%%%%%%%%%%%%%%%%%%%
%%%%%%%%%%%%%%%%%%%%%%%%%%%%%%%%%%%%%%%%%%%%%%%%%%%%%%%%%%%%%%%%%%%%%%%%%%%%%%%%%%%%%%%%%%%%%%%%%%%%%%%%%%%%%%%%%%%%%

The soft-wall model proposed in Ref.~\cite{Batell:2008zm} is a dynamical solution of Einstein's equation where in the string frame the metric is pure \abbr{AdS}.  It contains two scalar fields, loosely identified with the dilaton and tachyon of string theory.  In the Einstein frame the five-dimensional (5D) metric has the form
\begin{equation}
ds^2 = \EE^{-2 A(z)} \inp{ \eta_{\mu \nu} dx^\mu dx^\nu + dz^2 },
\label{Eq:background.lineelement.conformalcoord}
\end{equation}
with $\eta_{\mu\nu} = \diag{-1,+1,+1,+1}$ and a warp factor,
\begin{equation}
A(z) = \frac{2}{3} \inp{\mu z}^\nu + \ln \inp{k z},
\label{Eq:background.einsteinframe.metricfactor.conformalcoord}
\end{equation}
where $k$ is the \abbr{AdS} curvature scale.  Both scalar fields depend on the soft-wall mass scale $\mu$ and have a power-law profile in the conformal coordinate $z$ with an exponent that depends on the arbitrary parameter $\nu$.

The main interest here will be to perform a fluctuation analysis and obtain a Planck-weak scale hierarchy for \abbr{SM} fields interacting with the gravitational sector.  Although two scalar fields are required to obtain pure power-like dependence for the dilaton, they are not required to address the Planck-weak scale hierarchy: the scale $\mu$ will still set the weak scale regardless of the number of scalar fields.  We will therefore simplify our analysis by considering only a single real scalar field in the geometry \eq{background.einsteinframe.metricfactor.conformalcoord}.  The action is then
\begin{equation}
S_{\text{GRAV}}
	=	  M_5^3 \int_{z_0}^{\infty} \!\! dz \! \intOp[4]{x} \sqrt{-g} \,
			\inb{	  R
				- g^{MN} \pD_M \eta \pD_N \eta
				- V(\eta)
				}
		- M_5^3 \intOp[4]{x} \sqrt{-\gamma} \, \inb{ 2 K + \lambda_{\text{UV}}(\eta) },
\label{Eq:action.einsteinframe.gravity}
\end{equation}
with $M_5$ the $5D$ fundamental scale, $R$ the Ricci scalar, $g$ the determinant of the $5D$ metric (\ref{Eq:background.lineelement.conformalcoord}), $\gamma$ the determinant of the induced metric on the \abbr{UV} brane, $K$ the extrinsic curvature~\cite{York:1986fk}, and $\lambda_{UV}$ a \abbr{UV} boundary potential.  In this coordinate system the equations of motion are\footnote{It is outlined in \app{background} how to obtain these equations in the $y$-coordinate.  They may then be transformed to the $z$-coordinate using \eq{y.z.relation} to obtain the stated results.}
\begin{align}
3 A^{\prime\prime}(z) - 3 \bigl( A^\prime(z) \bigr)^2
	& =	\half (\pD_z \vev{\eta})^2 + \half \EE^{-2 A(z)} V(\vev{\eta}),
\label{Eq:einstein.eom.munu.conformalcoord}
	\\
6 \bigl( A^\prime(z) \bigr)^2
	& =	\half (\pD_z \vev{\eta})^2 - \half \EE^{-2 A(z)} V(\vev{\eta}),
\label{Eq:einstein.eom.zz.conformalcoord}
	\\
\EE^{-2 A(z)} \pderiv{V}{\vev{\eta}}
	& =	2 \pD_z^2 \vev{\eta} - 6 A^\prime(z) \pD_z \vev{\eta},
\label{Eq:scalar.eom.conformalcoord}
\end{align}
with boundary conditions
\begin{align}
0	& =	\inb[7mm]{ 6 A'(z) - \EE^{-A(z)} \lambda_{UV}(\vev{\eta}) }_{z = z_0},
\label{Eq:metric.BC.conformalcoord}
	\\
0	& =	\inb[7mm]{ 2 \pD_z \vev{\eta} - \EE^{-A(z)} \pderiv{\lambda_{UV}}{\eta} }_{z = z_0},
\label{Eq:scalar.BC.conformalcoord}
\end{align}
and where the classical background solution, denoted by $\vev\eta$, is assumed to be only a function of the coordinate $z$.  There are three integration constants from the equations of motion \eqn{einstein.eom.munu.conformalcoord}--\eqn{scalar.eom.conformalcoord}: $\eta(z_0), \eta^\prime(z_0)$, and $A(z_0)$.  The constant $A(z_0)$ is completely irrelevant because it does not affect
the equations of motion or the boundary conditions.  This would seem to imply that the two boundary conditions \eqn{metric.BC.conformalcoord} and \eqn{scalar.BC.conformalcoord} are then sufficient to determine the two remaining integration constants; however---as we will show in the next subsection---the metric is in fact singular and there is an extra constraint at the singularity.  This means there is a hidden fine-tuning which is identified with tuning the cosmological constant to zero~\cite{Forste:2000ft,Forste:2000ps,Cabrer:2009we}.

The solutions of equations \eqn{einstein.eom.munu.conformalcoord}--\eqn{scalar.BC.conformalcoord} are most simply obtained by using the superpotential method~\cite{DeWolfe:1999cp, Skenderis:1999mm}.  This gives rise to the solutions
\begin{align}
\vev{\eta}
 	& = \pm \sqrt{3} \inp{\frac{\nu + 1}{\nu}} \inb{
							  \sqrt{ \frac{2}{3} \frac{\nu}{\nu + 1} \inp{\mu z}^\nu + \Bigl(\frac{2}{3} \frac{\nu}{\nu + 1} \inp{\mu z}^\nu \Bigr)^2 }
							+ \sinh^{-1}\inp{    \sqrt{ \frac{2}{3} \frac{\nu}{\nu + 1} \inp{\mu z}^\nu }    }
							},
\label{Eq:scalar.vev.conformalcoord}
\\
V(z)	& =	- k^2 \EE^{\frac{4}{3} \inp{\mu z}^\nu} \left[  12 + 2 \nu \inp{7 - \nu} \inp{\mu z}^\nu + 4 \nu^2 \inp{\mu z}^{2\nu}  \right],
\label{Eq:potential.bulk.func.of.conformalcoord}
	\\
\lambda_{UV}
	& = 	  k\,\EE^{\frac{2}{3}\inp{ \mu z_0 }^\nu} 
			\biggl[
				6 + 4 \nu  \inp{\mu z_0}^\nu
			\biggr.
\notag	\\
	& \hphantom{= k \EE^{2 \inp{ \mu z_0 }^\nu / 3} \biggl[6\biggr.} \;
			\biggl. {}
				+ 2 \sqrt{3} \inp{1 + \nu} \sqrt{\frac{2}{3} \frac{\nu}{\nu + 1} \inp{\mu z_0}^\nu + \Bigl( \frac{2}{3} \frac{\nu}{1 +\nu} (\mu z_0)^\nu \Bigr)^2 } \,
					\bigl( \vev{\eta} - \eta_0 \bigr)
			\biggr]
		+ \dotsb
\label{Eq:potential.boundary.func.of.conformalcoord},
\end{align}
where $\eta_0$ is an arbitrary constant.  The boundary conditions are satisfied provided that $\vev{\eta}_{z_0} = \eta_0$. 

The expression for $\vev{\eta}$ simplifies in various regimes: for $z \sim 0$,
\begin{equation}
\lim_{z \to 0} \vev{\eta} \sim 2 \sqrt{2} \, \sqrt{ \frac{1 + \nu}{\nu} } \inp{\mu z}^{\nu/2}
\label{Eq:scalar.vev.z.near.zero},
\end{equation}
corresponding to the ``tachyon" of~\cite{Batell:2008zm}.  Meanwhile for large $z$,
\begin{equation}
\lim_{z \to \infty} \vev{\eta} \sim \frac{2}{\sqrt{3}} \inp{\mu z}^{\nu}.
\label{Eq:scalar.vev.z.near.infinity}
\end{equation}
Thus, in the far \abbr{IR} $\vev{\eta}$ is like the dilaton of~\cite{Batell:2008zm}; however, unlike like Ref.~\cite{Batell:2008zm}, the metric is no longer pure \abbr{AdS} in the string frame.

%%%%%%%%%%%%%%%%%%%%%%%%%%%%%%%%%%%%%%%%%%%%%%%%%%%%%%%%%%%%%%%%%%%%%%%%%%%%%%%%%%%%%%%%%%%%%%%%%%%%%%%%%%%%%%%%%%%%%
% % % % % % % % % % % % % % % % % % % % % % % % % % % % % % % % % % % % % % % % % % % % % % % % % % % % % % % % % % %
\subsection{Curvature singularity}
% % % % % % % % % % % % % % % % % % % % % % % % % % % % % % % % % % % % % % % % % % % % % % % % % % % % % % % % % % %
%%%%%%%%%%%%%%%%%%%%%%%%%%%%%%%%%%%%%%%%%%%%%%%%%%%%%%%%%%%%%%%%%%%%%%%%%%%%%%%%%%%%%%%%%%%%%%%%%%%%%%%%%%%%%%%%%%%%%

Although the conformally flat coordinate $z$ elucidates many features of the solution, it obscures an important one---namely that there is a naked singularity resulting in the space being finite~\cite{Gubser:2000nd,Cabrer:2009we}.  In the $z$ coordinate the singularity in the scalar curvature,
\begin{equation}
R(z) = - \frac{4}{3} k^2 \EE^{ \frac{4}{3} \inp{\mu z}^\nu } \left[ 4 \nu^2 \inp{\mu z}^{2 \nu} + 4 \nu \inp{4 - \nu} \inp{\mu z}^\nu + 15 \right],
\label{Eq:scalar.curvature.conformalcoord}
\end{equation}
occurs only at infinity; however, a coordinate redefinition\footnote{The papers ~\cite{Batell:2008zm,Batell:2008me} use the exponential integral, $-\Ei(-x)$, instead of the incomplete gamma function, $\Gamma(0,x)$.  For $x > 0$ we have $-\Ei(-x) = \Gamma(0,x)$.},
\begin{equation}
y(z) = \frac{1}{k \nu} \inb[0.75cm]{  \Gamma\inp{0, {\textstyle \frac{2}{3}} (\mu z_0)^\nu } - \Gamma\inp{0, {\textstyle \frac{2}{3}} \inp{\mu z}^\nu }  },
\label{Eq:y.z.relation}
\end{equation}
with resulting line element
\begin{equation}
ds^2 = \EE^{-2 A(y)} \eta_{\mu \nu} dx^\mu dx^\nu + dy^2,
\label{Eq:background.lineelement.ycoord}
\end{equation}
reveals that the space is indeed finite:
\begin{equation}
y(\infty) \equiv y_s = \frac{1}{k \nu} \Gamma\inp{0, {\textstyle \frac{2}{3}} \inp{\mu z_0}^\nu }\simeq -\frac{1}{k}\log(\mu z_0).
\label{Eq:ys.definition}
\end{equation}

The presence of the naked singularity implies that spacetime ends at $y = y_s$ and it must be checked that the boundary terms of the equations of motion vanish there (lest a non-zero 4D cosmological constant be generated).  To ensure this self-consistency of the theory we need to determine the superpotential as a function of the scalar field $\eta$, which is most simply done using the $y$ coordinate.  This requires obtaining $z(y)$ by utilizing \eq{ys.definition} along with \eq{y.z.relation} to yield
\begin{equation}
\Gamma^{-1}\bigl(0, k \nu \inp{y_s - y} \bigr) = \frac{2}{3} \inp{\mu z}^\nu
\label{Eq:inversefunc.definition}.
\end{equation}
The metric factor then becomes
\begin{equation}
A(y) = \Gamma^{-1}\bigl(0, k \nu \inp{y_s - y}\bigr) + \frac{1}{\nu} \ln\inb{  \Gamma^{-1}\bigl(0, k \nu \inp{y_s - y} \bigr)  } + \frac{1}{\nu} \ln\inb{ \frac{3}{2} \inp{\frac{k}{\mu}}^\nu }.
\label{Eq:background.einsteinframe.metricfactor.ycoord}
\end{equation}
Using the differentiation rule for $\Gamma^{-1}$
\begin{equation}
\deriv{}{y} \Gamma^{-1}\bigl( k \nu (y_s - y) \bigr) = k \nu \Gamma^{-1}\bigl( k \nu (y_s - y) \bigr) \EE^{\Gamma^{-1}( k \nu (y_s - y) )}
\label{Eq:GammaZeroInverse.differentiation.rule},
\end{equation}
along with employing the superpotential technique~\cite{Skenderis:1999mm,DeWolfe:1999cp}, the equations of motion (see \app{background} for the equations in the $y$ coordinate) may be readily solved.  The result for $\vev{\eta}$ is just \eq{scalar.vev.conformalcoord} written in the $y$ coordinate,
\begin{equation}
\vev{\eta}
 	= \pm \sqrt{3} \inp{\frac{\nu + 1}{\nu}} \inb{
		\sqrt{ \frac{\nu}{\nu + 1} \Gamma^{-1} + \Bigl(\frac{\nu}{\nu + 1} \Gamma^{-1}\Bigr)^2 }
							+ \sinh^{-1}\inp{ \sqrt{\frac{\nu}{\nu + 1} \Gamma^{-1}} }
						}.
\label{Eq:scalar.vev}
\end{equation}
This expression, along with metric factor \eq{background.einsteinframe.metricfactor.ycoord}, may be simplified in the \abbr{UV},
\begin{align}
\left.\vev{\eta}\right|_{y \sim 0}
	& \simeq		2 \sqrt{2} \, \sqrt{ \frac{1 + \nu}{\nu} } \inp{\frac{\mu}{k}}^{\nu/2} \EE^{\frac{\nu}{2} k y},
\label{Eq:scalar.vev.y.near.zero}
	\\
\left.A(y)\right|_{y \sim 0}
	& \simeq		k y + \frac{2}{3} \inp{ \frac{\mu}{k} }^\nu \EE^{\nu k y},
\label{Eq:background.einsteinframe.metricfactor.ycoord.near.zero}
\end{align}
as well as the \abbr{IR},
\begin{align}
\left.\vev{\eta}\right|_{y \sim y_s}
	& \simeq		- \sqrt{3} \ln \inp[3ex]{ k \nu (y_s - y) },
\label{Eq:scalar.vev.y.near.ys}
	\\
\left.A(y)\right|_{y \sim y_s}
	& \simeq 	- \ln \inp[3ex]{k \nu (y_s - y) }
\label{Eq:background.einsteinframe.metricfactor.ycoord.near.zys}.
\end{align}
Near $y = 0$ the scalar field \eq{scalar.vev.y.near.zero} is equivalent to the scalar field of the explicit model considered in Section 4 of~\cite{DeWolfe:1999cp} (and also reviewed in \app{DFGKModel}).
Comparison of \eq{scalar.vev.y.near.ys} with the equivalent expression in \cite{Cabrer:2009we} shows that our geometry matches that of \cite{Cabrer:2009we} at a single value $\nu=1$, and represents a different class of solutions for $\nu\neq 1$. 

The superpotential is formally obtained by inverting \eq{scalar.vev} to get $y(\vev{\eta})$ and substituting this into \eq{metric.superW.relation.ycoord} to produce $W$.  Note that \eq{scalar.vev} 
cannot be inverted analytically; however, what can be done is to introduce a parameter $\beta$ given by
\begin{equation}
\sinh \, \vev{\beta}
	= \sqrt{\frac{\nu}{\nu + 1} \Gamma^{-1}}
\label{Eq:definition.beta},
\end{equation}
write \eq{scalar.vev} in terms of $\beta$, and then solve the superpotential relationship with respect to $\beta$.  Doing so leads to an implicitly defined superpotential, $W(\beta(\eta))$, given by 
\begin{align}
W	& =	\half k \nu \inb{ \frac{\nu + 1}{2 \nu} \inp{1 + \cosh 2\beta} - 1 } \EE^{ \inp{\nu + 1}\inp{\cosh 2\beta - 1}/2 \nu },
\label{Eq:superW.func.beta}
	\\
\eta
	& =	\pm \sqrt{3} \, \frac{\nu + 1}{\nu} \inp{ \half \sinh 2\beta + \beta },
\label{Eq:scalar.func.beta}
\end{align}
with the latter equation implicitly defining $\beta(\eta)$.

Although \eq{scalar.func.beta} cannot be inverted to get the superpotential as an explicit function of $\eta$, 
it can be simplified in asymptotic regimes.  For large $\eta$, and consequently large $\beta$, we have
\begin{equation}
\lim_{\eta \to \infty} W \simeq \Bigl( \frac{\eta}{\sqrt{3}} \Bigr)^{ \!\! \frac{\nu - 1}{2 \nu} } \EE^{\pm \frac{\eta}{\sqrt{3}}}
\label{Eq:superW.asymptotic.form}.
\end{equation}
This regime is interesting because it is precisely the regime that determines whether the induced boundary terms vanish at $y = y_s$ thus leaving a zero $4D$ cosmological constant.  The specific condition is that the superpotential grows asymptotically slower than $\EE^{\frac{2\eta}{\sqrt{3}}}$~\cite{Cabrer:2009we}, which is satisfied by \eq{superW.asymptotic.form}.  The net result of meeting this criterion is that there is no need to resolve the singularity; that is, knowledge of the \abbr{UV} completion is unnecessary as we have a consistent solution to Einstein's equations on $[0,y_s)$.

%%%%%%%%%%%%%%%%%%%%%%%%%%%%%%%%%%%%%%%%%%%%%%%%%%%%%%%%%%%%%%%%%%%%%%%%%%%%%%%%%%%%%%%%%%%%%%%%%%%%%%%%%%%%%%%%%%%%%
%%%%%%%%%%%%%%%%%%%%%%%%%%%%%%%%%%%%%%%%%%%%%%%%%%%%%%%%%%%%%%%%%%%%%%%%%%%%%%%%%%%%%%%%%%%%%%%%%%%%%%%%%%%%%%%%%%%%%
\section{Obtaining a Hierarchy}
\label{Sec:Hierarchy}
%%%%%%%%%%%%%%%%%%%%%%%%%%%%%%%%%%%%%%%%%%%%%%%%%%%%%%%%%%%%%%%%%%%%%%%%%%%%%%%%%%%%%%%%%%%%%%%%%%%%%%%%%%%%%%%%%%%%%
%%%%%%%%%%%%%%%%%%%%%%%%%%%%%%%%%%%%%%%%%%%%%%%%%%%%%%%%%%%%%%%%%%%%%%%%%%%%%%%%%%%%%%%%%%%%%%%%%%%%%%%%%%%%%%%%%%%%%

With the background solutions in hand, it is now possible to discuss getting a hierarchy of scales; that is, the possibility of obtaining the Planck-weak scale hierarchy.  To achieve such a hierarchy, it is necessary to examine the inputs to the theory.  Although we began by specifying the background metric, \eq{background.einsteinframe.metricfactor.conformalcoord}, this is not an input to our theory; rather, the physical theory is described by a bulk and boundary potential which then determine the background solutions, Eqs.~\eqn{background.einsteinframe.metricfactor.conformalcoord} and \eqn{scalar.vev.conformalcoord}.  The bulk potential is determined solely from the superpotential, \eq{superW.func.beta}, and therefore has only the parameters $k$ and $\nu$ since these are the only parameters found in $W$.  The boundary potential, 
\begin{equation}
\lambda_{\text{UV}}
	= 12 W\big|_{\eta_0} + 12 \pD_\eta W\big|_{\eta_0}\inp{\eta - \eta_0} + m_{\text{UV}} \inp{\eta - \eta_0}^2
\label{Eq:boundary.potential},
\end{equation}
is also determined by the superpotential, but it additionally introduces boundary parameters $\eta_0$ and $m_{\text{UV}}$.  The inputs to our theory are thus $k$, $\nu$, $\eta_0$, and $m_{\text{UV}}$ of which some combination should determine the weak scale, $\mu$.  

Enforcing the boundary conditions, Eqs.~\eqn{metric.BC.conformalcoord}--\eqn{scalar.BC.conformalcoord}, reveals that $\left.\vev{\eta}\right|_{z = z_0} = \eta_0$  so the relationship between $\mu$ and our input parameters is
\begin{align}
\eta_0
 	& =	\pm \sqrt{3} \inp{\frac{\nu + 1}{\nu}} \inb{
							  \sqrt{		  \frac{2}{3} \frac{\nu}{\nu + 1} \inp{ \frac{\mu}{k} }^\nu 
							  		+ \Bigl(\frac{2}{3} \frac{\nu}{\nu + 1} \inp{ \frac{\mu}{k} }^\nu \Bigr)^2
								}
							+ \sinh^{-1} \sqrt{    \frac{2}{3} \frac{\nu}{\nu + 1} \inp{ \frac{\mu}{k} }^\nu    }
						},
\label{Eq:scalar.zero.at.yiszero}
	\\
	& \simeq	
		2\sqrt{2} \sqrt{\frac{1+\nu}{\nu}} \left(\frac{\mu}{k}\right)^{\nu/2}
\label{Eq:scalar.zero.at.yiszero.approx}.
\end{align}
Using \eq{scalar.zero.at.yiszero.approx}, we then obtain $\mu$ as a function of $\eta_0$, $k$, and $\nu$:
\begin{equation}
\mu = k \inp{ \frac{\nu \eta_0^2}{8 (1 + \nu)} }^{1/\nu}
\label{mu.as.function.of.eta0},
\end{equation}
where it is seen that for $\eta_0 \sim 0.1$, $\nu \sim 1/8$ we can obtain the desired Planck-weak scale hierarchy.  It is relevant to ask whether these input values are natural; that is, to determine what range of values our inputs would be expected to take.  As $\eta_0$ is a \abbr{UV} boundary term, it is expected to be of order the \abbr{UV} scale or, given our units, it should be of order one.  To address the normal values of $\nu$ it is necessary to know how it enters the potential.  Later on we show that the mass of $\eta$ is dependent on $\nu k$ (see \eq{potential.expanded.about.zero}), and since this is a bulk mass it is expected that it should be of order $k$---hence $\nu$ should also be of order one.  

\begin{figure}[htb]
\begin{center}
\framebox{
\begin{picture}(310,210)
\put(0,0){\includegraphics{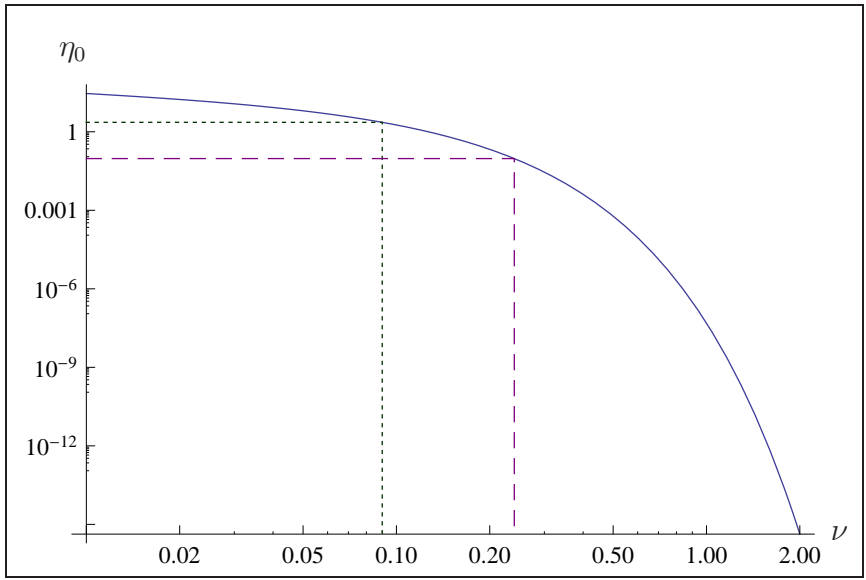}}
%
% vertical axis
\put(13,194){$\eta_0$}
%
% horizontal axis
\put(305,11){$\nu$}
\end{picture}
}
\end{center}
\caption{The value of the scalar field \abbr{VEV} at the \abbr{UV} brane $\eta_0$, as a function of $\nu$ for fixed weak scale, $(2/3)(\mu/k)^\nu = (10^{-16})^\nu$.  The dotted and dashed lines mark the natural range of $\eta_0$ which correspond to $\nu = 0.09$, $\eta_0 = 2.30$ and $\nu = 0.24$, $\eta_0 = 0.09$, respectively.}
\label{Fig:VEV.scalar.ycoord.at.yiszero}
\end{figure}

With the natural values of our parameters determined, we may now examine in detail how to set the weak scale, $\mu$.  \fig{VEV.scalar.ycoord.at.yiszero} shows the dependence of $\eta_0$ on $\nu$ after fixing the Planck-weak scale hierarchy.  In \fig{VEV.scalar.ycoord.at.yiszero} it can be seen that for $\nu = 2$ the required value of 
$\eta_0$ is $\mathcal{O}(10^{-16})$---thus the hierarchy would need to be put in by hand.  More notable in \fig{VEV.scalar.ycoord.at.yiszero} is the improvement in generating the Planck-weak scale hierarchy for $\nu \sim 1$: by simply assuming $\eta_0 = 5 \times 10^{-8}$, a numerical quantity nine orders of magnitude smaller is generated.  Of course to naturally generate the Planck-weak scale hierarchy the value of $\eta_0$ should be of order one; this in turn implies that $0.09 \leq \nu \leq 0.24$ and so only order one values are used to generate the large hierarchy.

After generating the correct Planck-weak scale hierarchy we can also check the sensitivity of the system as the parameters are varied.  For $\nu \sim 1/4$ a one percent change in the weak scale causes around a $0.1\%$ change in $\eta_0$ meaning the system is fairly robust.  Altering the weak scale by $1\%$ for $\nu$ in the range $1 \le \nu \le 3$ changes $\eta_0$ by one percent as well, so that range is also robust.  As $\nu$ becomes larger, however, the system becomes quite sensitive to a variation of the parameters.

The value of $\nu$ has consequences for the perturbations of the fields around the background as well as for all bulk fields.  These consequences already manifest themselves when looking for the \abbr{CFT} dual to the field $\eta$: a series expansion around $\eta = 0$ results in a scalar potential
\begin{equation}
V(\eta) = -12 k^2 - k^2 \nu \inp{ 1 - \frac{\nu}{8} } \eta^2 + \dotsb.
\label{Eq:potential.expanded.about.zero}
\end{equation}
The scalar thus has a mass
\begin{equation}
m_\eta^2 = - 2 k^2 \nu \inp{ 1 - \frac{\nu}{8} },
\label{Eq:mass.scalar}
\end{equation}
implying, by the \abbr{AdS}/\abbr{CFT} correspondence, that $\eta$ is dual to an operator with dimension
\begin{equation}
\Delta = 2 + \half \abs{4 - \nu}
\label{Eq:scalar.dual.operator.dimension}.
\end{equation}
For $\nu = 1$ this operator has dimension $7/2$ and it will be seen in \sect{Fluctuations} that this corresponds to a scenario of unparticles with a mass gap~\cite{Falkowski:2008yr}.  Consequently values in the range $0 < \nu < 1$, lead to operators of fractional dimension, $7/2 < \Delta < 4$, which equate to unparticles without a mass gap.  Note also that as $\nu\rightarrow \infty$ the operator dimension $\Delta\rightarrow\infty$ which reproduces the hard-wall limit where conformal symmetry is broken by an operator of infinite dimension. 

%%%%%%%%%%%%%%%%%%%%%%%%%%%%%%%%%%%%%%%%%%%%%%%%%%%%%%%%%%%%%%%%%%%%%%%%%%%%%%%%%%%%%%%%%%%%%%%%%%%%%%%%%%%%%%%%%%%%%
%%%%%%%%%%%%%%%%%%%%%%%%%%%%%%%%%%%%%%%%%%%%%%%%%%%%%%%%%%%%%%%%%%%%%%%%%%%%%%%%%%%%%%%%%%%%%%%%%%%%%%%%%%%%%%%%%%%%%
\section{Fluctuation analysis}
\label{Sec:Fluctuations}
%%%%%%%%%%%%%%%%%%%%%%%%%%%%%%%%%%%%%%%%%%%%%%%%%%%%%%%%%%%%%%%%%%%%%%%%%%%%%%%%%%%%%%%%%%%%%%%%%%%%%%%%%%%%%%%%%%%%%
%%%%%%%%%%%%%%%%%%%%%%%%%%%%%%%%%%%%%%%%%%%%%%%%%%%%%%%%%%%%%%%%%%%%%%%%%%%%%%%%%%%%%%%%%%%%%%%%%%%%%%%%%%%%%%%%%%%%%

With the hierarchy in hand the next question to be addressed is whether the theory is stable under perturbations; that is, it must be ensured that there are no $4D$ tachyons.  To evaluate the fluctuations around the background, we perturb both the metric,
\begin{align}
ds^2	& = \EE^{2(F - A(z))} \inb{ \bigl((1 - 2 F)\eta_{\mu \nu} + h_{\mu \nu}\bigr) dx^\mu dx^\nu + 2 A_\mu dx^\mu dz + dz^2 },
\label{Eq:background.lineelement.conformalcoord.perturbed.radionform}
	\\
	& \approx \EE^{-2A(z)} \inb{ (\eta_{\mu \nu} + h_{\mu \nu}) dx^\mu dx^\nu + 2 A_\mu dx^\mu dz + (1 + 2 F) dz^2 }
\label{Eq:background.lineelement.conformalcoord.perturbed},
\end{align}
and the scalar field $\eta$,
\begin{equation}
\eta = \vev{\eta} + \tilde{\eta}
\label{Eq:scalar.magnitude.fluctuation},
\end{equation}
retaining only the linearized perturbations.  Note that $h_{\mu\nu}$ in \eq{background.lineelement.conformalcoord.perturbed} contains tensor, vector and scalar modes that require separate decompositions for massive and massless modes.  The scalar $F$ is the graviscalar that in the absence of $\eta$ corresponds to the radion.

The general system has been studied in~\cite{Csaki:2000zn,Kofman:2004tk,Kiritsis:2006ua}, and~\cite{Batell:2008me} addresses the tensor modes for the soft-wall geometry, \eq{background.einsteinframe.metricfactor.conformalcoord}.  The vector modes described by $A_\mu$ are all eaten by the massive tensor modes, except for the zero mode.  This zero mode profile depends strictly on the metric~\cite{Kiritsis:2006ua},
\begin{equation}
f_{A}(z) = (k z)^3 \EE^{2 \inp{\mu z}^\nu },
\label{Eq:gravivector.profile}
\end{equation}
is non-normalizable and therefore absent from the theory.  All that remains is to address the scalar modes.

%%%%%%%%%%%%%%%%%%%%%%%%%%%%%%%%%%%%%%%%%%%%%%%%%%%%%%%%%%%%%%%%%%%%%%%%%%%%%%%%%%%%%%%%%%%%%%%%%%%%%%%%%%%%%%%%%%%%%
% % % % % % % % % % % % % % % % % % % % % % % % % % % % % % % % % % % % % % % % % % % % % % % % % % % % % % % % % % %
\subsection{Scalar (Radion) modes}
\label{Sec:Fluctuations.Scalar.Modes}
% % % % % % % % % % % % % % % % % % % % % % % % % % % % % % % % % % % % % % % % % % % % % % % % % % % % % % % % % % %
%%%%%%%%%%%%%%%%%%%%%%%%%%%%%%%%%%%%%%%%%%%%%%%%%%%%%%%%%%%%%%%%%%%%%%%%%%%%%%%%%%%%%%%%%%%%%%%%%%%%%%%%%%%%%%%%%%%%%

%%%%%%%%%%%%%%%%%%%%%%%%%%%%%%%%%%%%%%%%%%%%%%%%%%%%%%%%%%%%%%%%%%%%%%%%%%%%%%%%%%%%%%%%%%%%%%%%%%%%%%%%%%%%%%%%%%%%%
%%  %%  %%  %%  %%  %%  %%  %%  %%  %%  %%  %%  %%  %%  %%  %%  %%  %%  %%  %%  %%  %%  %%  %%  %%  %%  %%  %%  %%  %
\subsubsection{Massive modes}
\label{Sec:Fluctuations.Scalar.Modes.Massive}
%%  %%  %%  %%  %%  %%  %%  %%  %%  %%  %%  %%  %%  %%  %%  %%  %%  %%  %%  %%  %%  %%  %%  %%  %%  %%  %%  %%  %%  %
%%%%%%%%%%%%%%%%%%%%%%%%%%%%%%%%%%%%%%%%%%%%%%%%%%%%%%%%%%%%%%%%%%%%%%%%%%%%%%%%%%%%%%%%%%%%%%%%%%%%%%%%%%%%%%%%%%%%%

Given the background solutions Eqs.~\eqn{background.einsteinframe.metricfactor.conformalcoord} and \eqn{scalar.vev.conformalcoord}, the massive scalar modes have the dynamical equation~\cite{Kofman:2004tk}
\begin{equation}
\inp{ \pD_z^2 + m^2 - V_\chi } \chi = 0.
\label{Eq:eom.massive.modes}
\end{equation}
The dynamical variable $\chi$ is given by
\begin{equation}
\chi \equiv - \sqrt{2} \frac{ \EE^{- \frac{3}{2} A(z)} }{A^\prime(z)}  (\pD_z \vev{\eta}) \inp{ -\frac{F}{2}  +  A^\prime(z) \frac{\tilde{\eta}}{\pD_z \vev{\eta}} },
\label{Eq:definition.dynamical.variable.massive.modes}
\end{equation}
and represents the admixture of the graviscalar, $F$, with the bulk scalar fluctuation, $\tilde{\eta}$, that describes the radion.  Its ``Schr\"odinger potential'' is
\begin{multline}
V_\chi
	= 	\frac{1}{4 z^2 \inp{1 + x}^2 \inp{1 + \nu + x}^2}
		\Bigl[	  (1 + \nu)^2 (5 - \nu) (3 - \nu)
			+ 6 (1 + \nu)(14 + 3 \nu - 2 \nu^2 - \nu^3) x
		\Bigr.
	\\
			\shoveright{ {}
			+ (195 + 186 \nu + 15 \nu^2 - 10 \nu^3 + \nu^4) x^2
			+ 2 (120 + 83 \nu + 6 \nu^2 + \nu^3) x^3
			\hspace{-1.5em}
			}
	\\
		\Bigr. {}
			+ 3 (55 + 24 \nu + \nu^2) x^4
			+ 12 (5 + \nu) x^5
			+ 9 x^6
		\Bigl],
\label{Eq:schrodinger.potential.massive.modes}
\end{multline}
with
\begin{equation}
x \equiv \frac{2}{3} \nu (\mu z)^\nu
\label{Eq:definition.x}.
\end{equation}
For $\nu \le 1$, \eq{schrodinger.potential.massive.modes} is always positive definite, so there are no negative mass-squared terms and therefore no 4D tachyons.  When $\nu < 1$, $V_\chi \to 0$ as $z \to \infty$ resulting in a continuum of modes which come arbitrarily close to, but never reach, zero (recall that \eq{schrodinger.potential.massive.modes} describes only massive modes).  When $\nu = 1$ the Schr\"odinger potential tends to the constant $\mu^2$ in the far \abbr{IR} so that there is a continuum of modes above $\mu^2$.  The potential as $z\to\infty$
behaves as $V_\chi \sim \frac{9}{4} \frac{x^2}{z^2}$, becoming infinitely large for $\nu>1$ and leading to a discrete particle spectrum.

%%%%%%%%%%%%%%%%%%%%%%%%%%%%%%%%%%%%%%%%%%%%%%%%%%%%%%%%%%%%%%%%%%%%%%%%%%%%%%%%%%%%%%%%%%%%%%%%%%%%%%%%%%%%%%%%%%%%%
%%  %%  %%  %%  %%  %%  %%  %%  %%  %%  %%  %%  %%  %%  %%  %%  %%  %%  %%  %%  %%  %%  %%  %%  %%  %%  %%  %%  %%  %
\subsubsection{Massless modes}
\label{Sec:Fluctuations.Scalar.Modes.Massless}
%%  %%  %%  %%  %%  %%  %%  %%  %%  %%  %%  %%  %%  %%  %%  %%  %%  %%  %%  %%  %%  %%  %%  %%  %%  %%  %%  %%  %%  %
%%%%%%%%%%%%%%%%%%%%%%%%%%%%%%%%%%%%%%%%%%%%%%%%%%%%%%%%%%%%%%%%%%%%%%%%%%%%%%%%%%%%%%%%%%%%%%%%%%%%%%%%%%%%%%%%%%%%%

As for the zero modes, Ref.~\cite{Kiritsis:2006ua} provides a detailed decomposition using the ``light-cone'' basis.  This allows an unambiguous separation of the massless tensor, vector and scalar modes of the gravitational sector by avoiding expressions with inverse powers of $\Box$.  In this basis the scalar profile equations are
\begin{align}
\pD_z \zeta
	& = 0,
\label{Eq:equation.scalar.profile.zero.mode.gauge.invariant}
	\\
\pD_z \inp{ \EE^{-3 A(z)} \frac{\tilde{\eta}}{\pD_z \vev{\eta}} }
	& = - 2 \EE^{-3 A(z)} \zeta,
\label{Eq:equation.scalar.profile.zero.mode.non.gauge.invariant}
\end{align}
where $\zeta$ is defined as
\begin{equation}
\zeta \equiv - \frac{F}{2} + A^\prime(z) \frac{\tilde\eta}{\pD_z \vev{\eta}}.
\label{Eq:definition.scalar.gauge.invariant}
\end{equation}
These equations are readily solved, giving
\begin{align}
\zeta	& =	\zeta_0,
\label{Eq:scalar.profile.zero.mode.gauge.invariant}
	\\
\tilde{\eta}
	& =	\inb{2^{1+\frac{2}{\nu}} \frac{\mu^2}{\nu k^3} \zeta_0
		\inp{ \Gamma\Bigl(-\frac{2}{\nu}, 2(\mu z)^\nu\Bigr) - \Gamma\Bigl(-\frac{2}{\nu}, 2(\mu z_0)^\nu\Bigr) } + \frac{ \tilde{\eta}_0 \EE^{-3 A(z_0)} }{ \pD_z \vev{\eta}_0 }
		} \EE^{3 A(z)} \pD_z \vev{\eta},
\label{Eq:scalar.profile.zero.mode.non.gauge.invariant}
\end{align}
with $\zeta_0$ and $\tilde\eta_0$ constants.  The boundary conditions may be found in \cite{Csaki:2000zn,Kofman:2004tk}, and for simplicity we assume a large boundary mass term, $m_{\text{UV}} \to \infty$.  This results in the boundary conditions
\begin{align}
\left.\pD_z \inp{ \EE^{-2 A(z)} F }\right|_{z = z_0}	& = 0,
\label{Eq:BC.scalar.profile.zero.mode.gauge.invariant}
	\\
\left.\tilde\eta\right|_{z = z_0} 						& = 0.
\label{Eq:BC.scalar.profile.zero.mode.non.gauge.invariant}
\end{align}
The latter equation, \eq{BC.scalar.profile.zero.mode.non.gauge.invariant}, implies that $\tilde\eta_0 = 0$ resulting in both $\tilde\eta$ and $\zeta$ being proportional to $\zeta_0$.  Substituting the definition of $\zeta$, \eq{definition.scalar.gauge.invariant}, into the boundary condition \eq{BC.scalar.profile.zero.mode.gauge.invariant} yields the solution $\zeta_0 = 0$ and thus there are no zero modes.  We have checked that for arbitrary $m_{\text{UV}}$ the boundary conditions also demand that $\zeta_0 = \tilde\eta_0 = 0$ so there are no zero modes.

The fact that there are no massless modes can be intuitively seen from noting that the theory lacks the symmetry where the \abbr{UV} brane is shifted relative to the singularity---this being the analog of moving the UV and IR branes relative to each other in RS1.  The existence of this symmetry implies that there is a massless mode, and, as noted previously, this corresponds to the graviscalar $F$---also known as the radion.

We can now see why the radion is not massless for the soft wall: the scalar field in the soft wall is not invariant under a shift in the position of the \abbr{UV} brane since the boundary value $\eta_0 = \left.\vev{\eta}\right|_{y = y_{\text{UV}}}$ is explicitly fixed by the boundary conditions.  Thus, we cannot arbitrarily shift the position of the \abbr{UV} brane relative to the singularity, and therefore the radion is no longer massless.

Actually, because our soft-wall solution continuously varies between RS2 and RS1 a more exact relationship can be established.  For $\nu < 1$, the theory is RS2-like.  In this case there is no massless mode but the radion behaves like an unparticle or ``unradion".  Given that there is a continuum of positive momentum-squared, the effective $4D$ action of the unparticle at quadratic order will be positive definite and $\eta_0$ will be at a minimum.  In the limit $\nu\rightarrow 0$, the unparticle is no longer in the spectrum due to non-normalizability and we completely recover RS2.  When $\nu = 1$ the radion behaves like a continuum with a mass gap and the minimization will mimic the unHiggs scenario~\cite{Stancato:2008mp}.

For $\nu > 1$ the theory is RS1-like and the zero modes are present but acquire a non-zero mass.  For $\nu \gg 1$ the theory can be directly mapped to the RS1 case with a Goldberger-Wise scalar and an analytic expression for the radion mass can be obtained.  Using the DFGK model~\cite{DeWolfe:1999cp} (see \app{DFGKModel} for a brief review of the model and our notation) the correct relationship is
\begin{equation}
\frac{\nu}{2} = - \frac{u}{k},
\label{Eq:softwall.DFGK.relation}
\end{equation}
so that $\nu \to \infty$ corresponds to $u \to - \infty$.  In the large $\nu$ limit the physical regime is $z < 1/\mu$ and the scalar \abbr{VEV} is given by \eq{scalar.vev.z.near.zero} to good approximation.  The metric factor, \eq{background.einsteinframe.metricfactor.conformalcoord}, is
\begin{equation}
A(z) = \ln k z + \frac{16}{3} \vev{\eta}^2,
\label{Eq:softwall.largenu.metric}
\end{equation}
which compares favorably with \eq{DFGK.metricfactor} and since $z < 1/\mu$ the back reaction of the scalar is indeed negligible.

By employing the \abbr{WKB} approximation, the \abbr{KK} masses can be found~\cite{Batell:2008me},
\begin{equation}
m_n \simeq \mu \nu^{1/\nu} \pi^{1 - 1/\nu} n^{1 - 1/\nu},
\label{Eq:softwall.KK.masses}
\end{equation}
which agree with the DFGK model for $\mu = k \EE^{- \pi k R}$.  The radion mass, to order $\eta_0^2$, is then given by
\begin{equation}
m_{\text{radion}}^2 \simeq \third \nu^3 \mu^2 \inp{ \frac{\mu}{k} }^{\nu - 4},
\label{Eq:softwall.radion.mass}
\end{equation}
where $\nu > 4$ and as expected tends to zero as $\nu \to \infty$.  Thus we see that our soft wall solution asymptotes to RS1 with a massless radion and we obtain a light radion depending on the value $\eta_0 \sim (\mu/k)^{\nu/2}$.

%%%%%%%%%%%%%%%%%%%%%%%%%%%%%%%%%%%%%%%%%%%%%%%%%%%%%%%%%%%%%%%%%%%%%%%%%%%%%%%%%%%%%%%%%%%%%%%%%%%%%%%%%%%%%%%%%%%%%
%%%%%%%%%%%%%%%%%%%%%%%%%%%%%%%%%%%%%%%%%%%%%%%%%%%%%%%%%%%%%%%%%%%%%%%%%%%%%%%%%%%%%%%%%%%%%%%%%%%%%%%%%%%%%%%%%%%%%
\section{Bulk Fields in the Soft Wall}
\label{Sec:Bulk.Fields}
%%%%%%%%%%%%%%%%%%%%%%%%%%%%%%%%%%%%%%%%%%%%%%%%%%%%%%%%%%%%%%%%%%%%%%%%%%%%%%%%%%%%%%%%%%%%%%%%%%%%%%%%%%%%%%%%%%%%%
%%%%%%%%%%%%%%%%%%%%%%%%%%%%%%%%%%%%%%%%%%%%%%%%%%%%%%%%%%%%%%%%%%%%%%%%%%%%%%%%%%%%%%%%%%%%%%%%%%%%%%%%%%%%%%%%%%%%%

In order to make use of the Planck-weak scale hierarchy, it is necessary that the Higgs be a bulk 
field which naturally acquires a \abbr{VEV} of order $\mu$.  Since the Higgs is absorbed by the $W^{\pm}$ and $Z^0$ to make them massive, the gauge fields will then also need to be in the bulk.  For simplicity, however, we will assume that the fermions are confined to the \abbr{UV} brane.  The complete action for the bulk fields in the Einstein frame is then
\begin{equation}
S_E	= S_{\text{GRAV}} + S_{\text{GAUGE}} + S_{\text{HIGGS}},
\label{Eq:action.einsteinframe}
\end{equation}
with $S_{\text{GRAV}}$ defined by \eq{action.einsteinframe.gravity} and
\begin{align}
S_{\text{GAUGE}}
	& =	\intOp[5]{x} \sqrt{-g} \, \inb{ - \fourth \EE^{- \kappa(\eta)} g^{MR} g^{NS} F_{MN} F_{RS} },
\label{Eq:action.einsteinframe.U1.gauge}
	\\
S_{\text{HIGGS}}
	& =	  -\intOp[5]{x} \sqrt{-g} \,\EE^{3 \kappa(\eta)}
			\inb{	   g^{MN} \inp{\gD_M H}^\dagger \inp{\gD_N H}
				+ \EE^{4 \kappa(\eta)} V(H)
				}
		- \intOp[4]{x} \sqrt{-\gamma} \, V_{\text{UV}}(H),
\label{Eq:action.einsteinframe.scalar}
\end{align}
where for completeness a UV boundary potential, $V_{\text{UV}}(H)$, is also included for the Higgs.

In Eqs.~\eqn{action.einsteinframe.U1.gauge} and \eqn{action.einsteinframe.scalar}, $\kappa(\eta)$ is an arbitrary function of the scalar field $\eta$ and the exponential factors are chosen so that when $\kappa(\eta)= \frac{1}{2\sqrt{3}} \eta$ this corresponds to a canonical coupling which mimics the string theory dilaton in the string frame.  In general, however, the dilaton has a non-standard coupling to the \abbr[>>]{SM} bulk fields since $\kappa(\eta)$ is an arbitrary function.  We will see that arbitrary $\kappa$ leads to a variety of phenomenological possibilities and is crucial when $\nu < 1$ (when $\nu \geq 1$ there will always be at least a mass gap in the particle spectrum so arbitrary $\kappa$ is not critical).

%%%%%%%%%%%%%%%%%%%%%%%%%%%%%%%%%%%%%%%%%%%%%%%%%%%%%%%%%%%%%%%%%%%%%%%%%%%%%%%%%%%%%%%%%%%%%%%%%%%%%%%%%%%%%%%%%%%%%
% % % % % % % % % % % % % % % % % % % % % % % % % % % % % % % % % % % % % % % % % % % % % % % % % % % % % % % % % % %
\subsection{Gauge Fields}
\label{Sec:BulkFields.Gauge}
% % % % % % % % % % % % % % % % % % % % % % % % % % % % % % % % % % % % % % % % % % % % % % % % % % % % % % % % % % %
%%%%%%%%%%%%%%%%%%%%%%%%%%%%%%%%%%%%%%%%%%%%%%%%%%%%%%%%%%%%%%%%%%%%%%%%%%%%%%%%%%%%%%%%%%%%%%%%%%%%%%%%%%%%%%%%%%%%%

The equations of motion derived from the gauge field action \eq{action.einsteinframe.U1.gauge} can be transformed into a Schr\"odinger equation,
\begin{equation}
\inp{-\pD_z^2 + V_A(z)} \Psi_A	= -p^2 \Psi_A,
\label{Eq:schrodinger.gauge}
\end{equation}
with the substitution of the gauge boson profile $f_A = \EE^{\frac{1}{2} (A + \kappa)} \Psi_A$ 
and utilizing the gauge where $A_5 =0$ as well as $\eta^{\mu \nu} \pD_\mu A_\nu = 0$.  The potential for the 
soft-wall geometry, \eq{background.einsteinframe.metricfactor.conformalcoord}, is
\begin{align}
V_A(z)	& =	\frac{1}{4}(A' + \kappa')^2 -\frac{1}{2}(A''+\kappa''),
\notag	\\
	& =	  \frac{3}{4} \frac{1}{z^2}
		+ \frac{ \nu \inp{2 - \nu} }{3} \frac{ \mu^\nu }{ z^{2 - \nu} }
		+ \frac{\nu^2}{9} \frac{ \mu^{2\nu} }{ z^{2\inp{1 - \nu}} }
		+ \inp{ \frac{1}{2z} + \frac{\nu}{3} \frac{\mu^\nu}{z^{1 - \nu}} } \deriv{\kappa}{z}
		+ \fourth \inp{\deriv{\kappa}{z}}^2
		- \half \deriv[2]{\kappa}{z}.
\label{Eq:schrodinger.potential.stabilization.gauge}
\end{align}
Note that when $\kappa = 0$ the particle spectrum behavior is equivalent to that in the gravitational sector since all the $z$ dependence arises from $A(z)$.

We will focus on the regime where $\nu < 1$ since this gives rise to a Planck-weak scale hierarchy with natural values of the parameters.  For this $\nu$, then, the first three terms of \eq{schrodinger.potential.stabilization.gauge} tend to zero as $z \to \infty$ meaning the asymptotic behavior of $V_A$ is determined completely by the asymptotic behavior of $\kappa$:
\begin{list}{}{\setlength{\leftmargin}{9mm} \setlength{\rightmargin}{9mm} \setlength{\labelwidth}{0cm} \setlength{\itemsep}{0ex}}
\item[\textbf{Continuum:}]{When $\kappa(\eta) \sim \eta^{(1 - \epsilon)/\nu}$, corresponding to $\kappa \sim z^{1 - \epsilon}$, for $\epsilon > 0$, then large $z$ implies $V_A \to 0$.  The mass eigenstates $-p^2 = m^2$ are not quantized resulting in a continuous spectrum.}
\item[\textbf{Continuum with Mass Gap:}]{If $\kappa(\eta) \sim \eta^{1/\nu}$, resulting in $\kappa \sim b \mu z$ (where 
$b$ is an $\mathcal{O}(1)$ number), $V_A \to \fourth b^2 \mu^2$ as $z \to \infty$.  The mass spectrum is continuous above the mass gap $b \mu/2$.}
\item[\textbf{Discretum:}]{For $\kappa(\eta) \sim \eta^{(1 + \epsilon)/\nu}$ yielding the asymptotic behavior $\kappa \sim z^{1 + \epsilon}$ with $\epsilon > 0$, $V_A \to \infty$ as $z \to \infty$.  Since the potential goes to infinity, the mass modes are quantized and can be enumerated by an integer $n$.  A \abbr{WKB} approximation (using the asymptotic form for the potential) then obtains the eigenvalues $m_n^2 \propto \mu^2 n^{\frac{2}{1 + 1/\epsilon}}$.}
\end{list}

Thus we see that all types of particle spectrum behavior for the gauge fields can be obtained with the appropriate asymptotic behavior of $\kappa$.  Evidently for fields charged under the \abbr[>>]{SM}, a continuum without a mass gap is clearly not phenomenologically acceptable.  Therefore, for $\nu < 1$ a non-trivial $\kappa$ is necessary to provide these fields with either a mass gap or discretum.

%%%%%%%%%%%%%%%%%%%%%%%%%%%%%%%%%%%%%%%%%%%%%%%%%%%%%%%%%%%%%%%%%%%%%%%%%%%%%%%%%%%%%%%%%%%%%%%%%%%%%%%%%%%%%%%%%%%%%
% % % % % % % % % % % % % % % % % % % % % % % % % % % % % % % % % % % % % % % % % % % % % % % % % % % % % % % % % % %
\subsection{Scalar Fields}
\label{Sec:BulkFields.Scalar}
% % % % % % % % % % % % % % % % % % % % % % % % % % % % % % % % % % % % % % % % % % % % % % % % % % % % % % % % % % %
%%%%%%%%%%%%%%%%%%%%%%%%%%%%%%%%%%%%%%%%%%%%%%%%%%%%%%%%%%%%%%%%%%%%%%%%%%%%%%%%%%%%%%%%%%%%%%%%%%%%%%%%%%%%%%%%%%%%%

Just as for the gauge fields, the scalar field equations of motion may be transformed into a Schr\"odinger equation,
\begin{equation}
\inp{-\pD_z^2 + V_H(z)} \Psi_H = -p^2 \Psi_H,
\label{Eq:schrodinger.scalar}
\end{equation}
with the substitution of the Higgs field profile $f_H = \EE^{\frac{3}{2} (A - \kappa) } \Psi_H$.  Assuming a 5D potential, 
\begin{equation}
V(H) = M^2(z) H^\dagger H,
\label{Eq:potential.5D.higgs}
\end{equation}
with the Higgs mass, $M(z)$, a function of $z$, the Schr\"odinger potential becomes
\begin{align}
V_H(z)	& =	\frac{9}{4}(A'-\kappa')^2 -\frac{3}{2}(A''-\kappa'') + \EE^{4\kappa-2A} M^2,
\notag	\\
	& =	  \frac{15}{4} \frac{1}{z^2}
		+ \frac{ \nu \inp{4 - \nu} \mu^\nu }{ z^{2 - \nu} }
		+ \frac{\nu^2 \mu^{2\nu} }{ z^{2\inp{1 - \nu}} }
		- \inp{ \frac{9}{2z} + \frac{3 \nu \mu^\nu}{z^{1 - \nu}} } \deriv{\kappa}{z} + \frac{9}{4} \inp{\deriv{\kappa}{z}}^2
		+ \frac{3}{2} \deriv[2]{\kappa}{z}
\notag	\\
	& \quad {}
		+ \frac{1}{k^2 z^2} M^2(z) \EE^{4 \kappa(z) - \frac{4}{3} \inp{\mu z}^\nu }.
\label{Eq:schrodinger.potential.stabilization.higgs}
\end{align}
Once again for $\nu < 1$ the first three terms go to zero as  $z \to \infty$, leaving the asymptotic behavior dependent on 
$\kappa$ and $V(H)$.  If the contribution of the 5D potential is such that it tends to zero at large $z$ (for example, $2\ln M(z) + 4 \kappa(z)$ is asymptotically weaker than $\inp{\mu z}^\nu$), the behavior of this potential will parallel that of the gauge field; however, for a ``strong'' enough potential the behavior can be distinct from the gauge field.

An additional constraint associated with the Higgs is that since it acquires a \abbr{VEV} it could potentially back-react on the metric.  The energy-momentum tensor contribution of the Higgs is
\begin{align}
\vev{T_{55}}
	& =	\EE^{3 \kappa(\vev{\eta})} \inb{ \inp{\pD_z\vev{H}\mathclose{}}^2 - \EE^{4 \kappa(\vev{\eta})-2A(z)} M^2(z) \vev{H}^2 },
\label{Eq:energy.momentum.higgs.yy}
	\\
\vev{T_{\mu \nu}	}
	& \equiv		  \widehat{T} \eta_{\mu \nu}
	  =		- \half \eta_{\mu \nu} \EE^{3 \kappa(\vev{\eta})} \inb{ \inp{\pD_z\vev{H}\mathclose{}}^2 + \EE^{4 \kappa(\vev{\eta})-2 A(z)} M^2(z) \vev{H}^2 }
\label{Eq:energy.momentum.higgs.munu},
\end{align}
and for consistency this must be small compared to the contribution of $\eta$,
\begin{equation}
\vev{T_{55}} \sim \widehat{T} \sim M_5^4.
\label{Eq:energy.momentum.scalar}
\end{equation}
Evidently the back-reaction of the Higgs is dependent on the particular form of $\kappa(\eta)$, and in general it will be required that the \abbr{VEV} carry some exponential dependence to cancel the exponential dependence on $\kappa(\eta)$.  For $\nu > 1$ it is phenomenologically acceptable to take $\kappa(\eta) = \frac{1}{2\sqrt{3}}\eta$ (so that $\eta$ is proportional to the canonical dilaton).  This possibility was considered in \cite{Batell:2008me} where it was shown that the back-reaction due to the Higgs \abbr{VEV} goes as $k^4 < M_5^4$ so that it may indeed be neglected.  For arbitrary $\kappa(\eta)$ (which is required phenomenologically for $\nu < 1$) it is sufficient to require $\vev{H} \sim \EE^{-\frac{3}{2}\kappa}$ with $M^2(z) \sim \EE^{-4 \kappa}$ to cancel the exponential dependence.  This can be achieved, for example, by taking $\kappa(z) = (\mu z)^2 \sim \vev{\eta}^{2/\nu}$ and $M^2(z) = p(z) \EE^{- 4 (\mu z)^2 + \frac{4}{3}(\mu z)^\nu}$ with  $p(z)$ a polynomial in $z$.  The resulting Higgs \abbr{VEV} is then a Gaussian, $\vev{H} \sim \EE^{-a \inp{\mu z - 1}^2}$, for an arbitrary constant $a$ and centered around $z = 1/\mu$.  Thus, provided $a>3/2$, this example serves as a nice illustration that the Higgs back-reaction can be negligible even for $\nu < 1$ where arbitrary $\kappa$ is required.

%%%%%%%%%%%%%%%%%%%%%%%%%%%%%%%%%%%%%%%%%%%%%%%%%%%%%%%%%%%%%%%%%%%%%%%%%%%%%%%%%%%%%%%%%%%%%%%%%%%%%%%%%%%%%%%%%%%%%
%%%%%%%%%%%%%%%%%%%%%%%%%%%%%%%%%%%%%%%%%%%%%%%%%%%%%%%%%%%%%%%%%%%%%%%%%%%%%%%%%%%%%%%%%%%%%%%%%%%%%%%%%%%%%%%%%%%%%
\section{Conclusion}
\label{Sec:Conclusion}
%%%%%%%%%%%%%%%%%%%%%%%%%%%%%%%%%%%%%%%%%%%%%%%%%%%%%%%%%%%%%%%%%%%%%%%%%%%%%%%%%%%%%%%%%%%%%%%%%%%%%%%%%%%%%%%%%%%%%
%%%%%%%%%%%%%%%%%%%%%%%%%%%%%%%%%%%%%%%%%%%%%%%%%%%%%%%%%%%%%%%%%%%%%%%%%%%%%%%%%%%%%%%%%%%%%%%%%%%%%%%%%%%%%%%%%%%%%

We have examined the stabilization of a soft wall geometry based on the solution in \cite{Batell:2008zm}.  This geometry has the attractive features of being asymptotically \abbr{AdS} and having a scalar field that has an asymptotic power-law dependence in the conformal coordinate.  We found that the additional parameter of the power-law exponent permits a variety of mass spectra for the gravitational sector of the theory---ranging from a continuum without a mass gap to a discrete set of modes.  Correspondingly, we discovered that the exponent provided a means of continuously varying the theory between RS2 and RS1.  

We also examined obtaining a Planck-weak scale hierarchy in the model and discovered that a hierarchy could be generated with natural values of the parameters.  To ensure the stability of the hierarchy, a fluctuation analysis of the gravitational sector was performed where it was shown that there are no massless states and therefore the system is stabilized.  While the system is stabilized for arbitrary parameters, it was found that the natural solution to the hierarchy demands the gravitational sector have fields with a continuum of modes without a mass gap; specifically, in this regime, the radion acts like an unparticle or is an ``unradion''.  We also examined the limit where the theory approaches RS1 and made a direct comparison of our model to that of RS1 with a Goldberger-Wise scalar field.  We showed that the radion of our model is indeed massive and lighter than the soft wall mass scale.

Finally, we considered the addition of \abbr[>>]{SM} bulk fields in this background.  In the region of parameter space that naturally solves the Planck-weak scale hierarchy problem, we showed that these fields could obtain a continuum with a mass gap or a discrete mass spectrum provided nonstandard couplings to the gravity sector are permitted.  We investigated the implications of the nonstandard couplings with regards to the Higgs \abbr{VEV}'s back-reaction on the metric and demonstrated the existence of a solution where this back-reaction was negligible.  Given this self-consistent solution, we thus have the tantalizing possibility of \abbr{SM} fields interacting with a gravitational unparticle sector that helps to naturally generate the Planck-weak scale hierarchy.  Our soft wall model is therefore a more general way to address the hierarchy problem and provides a suitable starting point for a more detailed phenomenological analysis.

%%%%%%%%%%%%%%%%%%%%%%%%%%%%%%%%%%%%%%%%%%%%%%%%%%%%%%%%%%%%%%%%%%%%%%%%%%%%%%%%%%%%%%%%%%%%%%%%%%%%%%%%%%%%%%%%%%%%%
%%%%%%%%%%%%%%%%%%%%%%%%%%%%%%%%%%%%%%%%%%%%%%%%%%%%%%%%%%%%%%%%%%%%%%%%%%%%%%%%%%%%%%%%%%%%%%%%%%%%%%%%%%%%%%%%%%%%%
\section*{Acknowledgements}
\label{Sec:Acknowledgements}
%%%%%%%%%%%%%%%%%%%%%%%%%%%%%%%%%%%%%%%%%%%%%%%%%%%%%%%%%%%%%%%%%%%%%%%%%%%%%%%%%%%%%%%%%%%%%%%%%%%%%%%%%%%%%%%%%%%%%
%%%%%%%%%%%%%%%%%%%%%%%%%%%%%%%%%%%%%%%%%%%%%%%%%%%%%%%%%%%%%%%%%%%%%%%%%%%%%%%%%%%%%%%%%%%%%%%%%%%%%%%%%%%%%%%%%%%%%

We thank Brian Batell, Gero von Gersdorff, and Benedict von Harling for helpful discussions.  This work is supported by the Australian Research Council.

%%%%%%%%%%%%%%%%%%%%%%%%%%%%%%%%%%%%%%%%%%%%%%%%%%%%%%%%%%%%%%%%%%%%%%%%%%%%%%%%%%%%%%%%%%%%%%%%%%%%%%%%%%%%%%%%%%%%%
%   |   %   |   %   |   %   |   %   |   %   |   %   |   %   |   %   |   %   |   %   |   %   |   %   |   %   |   %   |
\appendix
%   |   %   |   %   |   %   |   %   |   %   |   %   |   %   |   %   |   %   |   %   |   %   |   %   |   %   |   %   |
%%%%%%%%%%%%%%%%%%%%%%%%%%%%%%%%%%%%%%%%%%%%%%%%%%%%%%%%%%%%%%%%%%%%%%%%%%%%%%%%%%%%%%%%%%%%%%%%%%%%%%%%%%%%%%%%%%%%%

%%%%%%%%%%%%%%%%%%%%%%%%%%%%%%%%%%%%%%%%%%%%%%%%%%%%%%%%%%%%%%%%%%%%%%%%%%%%%%%%%%%%%%%%%%%%%%%%%%%%%%%%%%%%%%%%%%%%%
%%%%%%%%%%%%%%%%%%%%%%%%%%%%%%%%%%%%%%%%%%%%%%%%%%%%%%%%%%%%%%%%%%%%%%%%%%%%%%%%%%%%%%%%%%%%%%%%%%%%%%%%%%%%%%%%%%%%%
\section{Background Equations and Solutions}
\label{App:background}
%%%%%%%%%%%%%%%%%%%%%%%%%%%%%%%%%%%%%%%%%%%%%%%%%%%%%%%%%%%%%%%%%%%%%%%%%%%%%%%%%%%%%%%%%%%%%%%%%%%%%%%%%%%%%%%%%%%%%
%%%%%%%%%%%%%%%%%%%%%%%%%%%%%%%%%%%%%%%%%%%%%%%%%%%%%%%%%%%%%%%%%%%%%%%%%%%%%%%%%%%%%%%%%%%%%%%%%%%%%%%%%%%%%%%%%%%%%

The action, \eq{action.einsteinframe.gravity}, in the $y$-coordinate defined by the line element \eq{background.lineelement.ycoord} is
\begin{equation}
S_{\text{GRAV}}
	=	  M_5^3 \int_{0}^{y_s} \!\! dy \! \intOp[4]{x} \sqrt{-g} \,
			\inb{	  R
				- g^{MN} \pD_M \eta \pD_N \eta
				- V(\eta)
				}
		- M_5^3 \intOp[4]{x} \sqrt{-\gamma} \, \inp{ 2 K + \lambda_{\text{UV}}(\eta) },
\label{Eq:action.einsteinframe.gravity.ycoord}
\end{equation}
with the terms defined in the text after \eq{action.einsteinframe.gravity}.  The extrinsic curvature $K$ is defined as \cite{York:1986fk}
\begin{align}
K	& =	\gamma^{MN} K_{MN},
\label{Eq:extrinsic.curvature}
	\\
K_{MN}	& =	- \half \inb{ \gamma_{MR} \pD_N n^R + \gamma_{RN} \pD_M n^R + n^R \pD_R \gamma_{MN} },
\label{Eq:extrinsic.curvature.tensor}
\end{align}
with the induced metric,
\begin{equation}
\gamma_{MN}
	=	g_{MN} - n_M n_N,
\label{Eq:induced.metric}
\end{equation}
where $n^M$ is a unit, outward-pointing normal vector to the boundary; in the $y$-coordinate $n^M = - \delta^M_y$.

By writing the action as in \eq{action.einsteinframe.gravity.ycoord}, we have separated the bulk gravitational physics from the boundary gravitational physics; that is, varying the first term of the action, \eq{action.einsteinframe.gravity.ycoord}, with respect to $g_{MN}$ (while ignoring the boundary) yields the bulk gravitational equations of motion
\begin{align}
\delta g^{\mu \nu}	&:&	 3 A''(y) - 6 \inp[5mm]{A'(y)}^2	 &=	\half (\pD_y \vev{\eta})^2  + \half V(\eta),
\label{Eq:einstein.eom.munu}
	\\
\delta g^{55} &:& 6 \inp[5mm]{A'(y)}^2			 &=	\half (\pD_y \vev{\eta})^2 - \half V(\eta),
\label{Eq:einstein.eom.yy}
\end{align}
whilst varying the second term of \eq{action.einsteinframe.gravity.ycoord} with respect to $\gamma_{MN}$ yields the gravitational boundary condition
\begin{align}
0	& =	\inb[7mm]{ 6 A'(y) - \lambda_{UV}(\vev{\eta}) }_{y = 0}.
\label{Eq:metric.BC}
\end{align}
Varying the action with respect to $\eta$ yields the scalar equation of motion and boundary condition,
\begin{align}
\pderiv{V}{\eta}		& =	2 \pD_y^2 \vev{\eta} - 8 A'(y) \pD_y \vev{\eta},
\label{Eq:scalar.eom}
\intertext{and}
0			& =	\inb[7mm]{ 2 \pD_y \vev{\eta} - \pderiv{\lambda_{UV}}{\eta} }_{y = 0},
\label{Eq:scalar.BC}
\end{align}
respectively.

The superpotential method utilizes the assignment
\begin{align}
A'(y)	& = 	2 W,
\label{Eq:metric.superW.relation.ycoord}
	\\
\pD_y \vev{\eta}
	& =	6 \pderiv{W}{\eta},
\label{Eq:scalar.superW.relation.ycoord}
\end{align}
with the bulk and boundary potentials
\begin{align}
V(\eta)
	& =	  36 \inp{\pderiv{W}{\eta}}^2 - 48 W^2,
	\\
\lambda_{UV}(\eta)
	& = 	  12 \left.W\right|_{\eta = \eta_0}
		+ 12 \left.\pderiv{W}{\eta}\right|_{\eta = \eta_0} \inp{ \eta - \eta_0 }
		+ m_{\text{UV}} \inp{ \eta - \eta_0 }^2.
\end{align}

%%%%%%%%%%%%%%%%%%%%%%%%%%%%%%%%%%%%%%%%%%%%%%%%%%%%%%%%%%%%%%%%%%%%%%%%%%%%%%%%%%%%%%%%%%%%%%%%%%%%%%%%%%%%%%%%%%%%%
%%%%%%%%%%%%%%%%%%%%%%%%%%%%%%%%%%%%%%%%%%%%%%%%%%%%%%%%%%%%%%%%%%%%%%%%%%%%%%%%%%%%%%%%%%%%%%%%%%%%%%%%%%%%%%%%%%%%%
\section{DFGK Model}
\label{App:DFGKModel}
%%%%%%%%%%%%%%%%%%%%%%%%%%%%%%%%%%%%%%%%%%%%%%%%%%%%%%%%%%%%%%%%%%%%%%%%%%%%%%%%%%%%%%%%%%%%%%%%%%%%%%%%%%%%%%%%%%%%%
%%%%%%%%%%%%%%%%%%%%%%%%%%%%%%%%%%%%%%%%%%%%%%%%%%%%%%%%%%%%%%%%%%%%%%%%%%%%%%%%%%%%%%%%%%%%%%%%%%%%%%%%%%%%%%%%%%%%%

The DFGK model is useful for exploring a Goldberger-Wise scalar in RS1 and has been thoroughly examined in the literature~\cite{DeWolfe:1999cp,Csaki:2000zn,Kofman:2004tk}.  A brief overview of the model is given here.  The superpotential for the DFGK model is
\begin{equation}
W = 12 k - u \phi^2,
\label{Eq:DFGK.superW}
\end{equation}
where the action for the $\phi$ is the same as \eq{action.einsteinframe.gravity} with $\eta = \phi/\sqrt{2}$.  The resulting bulk and brane potentials are
\begin{align}
V	& =	-12 k^2 + \half (4 k u + u^2) \phi^2 - \frac{u^2}{12} \phi^4,
\label{Eq:DFGK.bulk.potential}
	\\
\lambda_{\text{UV}}
	& =	12 k - u \phi_0^2 - 2 u \phi_0 (\phi - \phi_0) + \half m_{\text{UV}} (\phi - \phi_0)^2,
\label{Eq:DFGK.UV.potential}
\end{align}
and the \abbr{IR} potential given by $\lambda_{\text{UV}}$ with $k \to -k$, $u \to - u$, $\phi_0 \to \phi_{\text{IR}}$, and $m_{\text{UV}} \to m_{\text{IR}}$.  The background solutions in the $y$-coordinate are
\begin{align}
\vev{\phi} 
	& =	\phi_0 \EE^{- u y},
\label{Eq:DFGK.scalar.vev}
	\\
A(y)	& =	k y + \frac{1}{24} \inp{\! \vev{\phi}^2 - \phi_0^2 }.
\label{Eq:DFGK.metricfactor}
\end{align}
For $\phi_0 \ll 1$ the geometry is essentially \abbr{AdS} between the branes and the \abbr{KK} masses are
\begin{equation}
m_n \approx n \pi k \EE^{- \pi k R}.
\label{Eq:DFGK.KK.masses}
\end{equation}
The radion mass, to leading order in $\phi_0$, is
\begin{equation}
m^2_{\text{radion}} \approx - \frac{	\phi_0^2 u^2 (2k + u) \EE^{- 2 \pi u R} \inp{\EE^{2 k \pi R} - 1}	}
				{	3 k \inp{ \EE^{-2 \pi u R} - \EE^{4 \pi k R} }				}.
\label{Eq:DFGK.radion.mass}
\end{equation}

% <references>
\bibliography{%
	ads-cft,%
	conformal_field_theory,%
	extradimensions,%
	gravity-classical,%
	xdim-mass_hierarchy,%
	flavor_violation,%
	unparticle%
}
% </references>

\end{document}